\documentclass[journal]{IEEEtran}
\IEEEoverridecommandlockouts
\usepackage{amsmath}
\usepackage{mathrsfs,amsthm,amsmath,amssymb}
\usepackage{textcomp}
\usepackage{graphicx,multirow}
\usepackage{balance}
\usepackage{cite}
\usepackage{gensymb}
\usepackage{algorithm}
\usepackage{algorithmic}
\usepackage[center]{caption}

\usepackage{xcolor}

\newcommand{\bs}{\boldsymbol}
\newcommand{\mb}{\mathbf}

\DeclareMathOperator{\st}{s.t.}

\begin{document}

\title{\textcolor{black}{Energy-efficient Deep Reinforcement Learning-based
Network Function Disaggregation in Hybrid Non-terrestrial Open
Radio Access Networks} }
\author{S. M. Mahdi Shahabi, Xiaonan Deng, Ahmad Qidan, Taisir Elgorashi and Jaafar Elmirghani

\thanks{S. M. Mahdi Shahabi, Xiaonan Deng, Ahmad Qidan, Taisir Elgorashi and Jaafar Elmirghani are with the Department of Engineering, King's College London, U.K. (e-mail: mahdi.shahabi@kcl.ac.uk;  xiaonan.deng@kcl.ac.uk; ahmad.qidan@kcl.ac.uk; taisir.elgorashi@kcl.ac.uk; jaafar.elmirghani@kcl.ac.uk).}
}
 \markboth{}{}
 \IEEEpubid{}

\maketitle

\begin{abstract}
This paper explores the integration of Open Radio Access Network (O-RAN) principles with non-terrestrial networks (NTN) and investigates the optimization of the functional split between Centralized Units (CU) and Distributed Units (DU) to improve energy efficiency in dynamic network environments. Given the inherent constraints of NTN platforms, such as Low Earth Orbit (LEO) satellites and high-altitude platform stations (HAPS), we propose a reinforcement learning-based framework utilizing Deep Q-Network (DQN) to intelligently determine the optimal RAN functional split. The proposed approach dynamically adapts to real-time fluctuations in traffic demand, network conditions, and power limitations, ensuring efficient resource allocation and enhanced system performance. 
The numerical results demonstrate that the proposed policy effectively adapts to network traffic flow by selecting an efficient network disaggregation strategy and corresponding functional split option based on data rate and latency requirements. 
\end{abstract}


\section{Introduction}\label{Introduction}


Non-Terrestrial Networks (NTNs), which include satellite systems, High-Altitude Platform Stations (HAPS), and aerial communication nodes, have emerged as vital enablers in extending network coverage to underserved, remote, and mobile regions where conventional terrestrial infrastructure is either impractical or cost-prohibitive. As NTNs continue to evolve and become a core component of next-generation communication systems, particularly within the scope of 6G, their integration with emerging network paradigms is increasingly critical to achieving global, seamless, and resilient connectivity. In parallel with advancements in NTNs, the emergence of Open Radio Access Networks (O-RAN) is transforming conventional network architectures. O-RAN introduces a paradigm shift by promoting an open, modular, and software-driven network infrastructure, which harnesses key enablers such as disaggregation, virtualization, and programmability. This architectural framework significantly departs from traditional, monolithic hardware-based approaches, allowing network components like Radio Units (RU), Distributed Units (DU), and Centralized Units (CU) to be deployed with increased flexibility and scalability. By decoupling hardware and software, O-RAN enables network operators to implement adaptive, software-defined functionalities, making it easier to accommodate shifting network conditions and varying traffic demands in real time. The benefits of integrating O-RAN with NTNs are numerous, offering improved efficiency and adaptability; however, this convergence also brings forth significant challenges, particularly concerning the efficient management of energy resources.

In terrestrial networks, power constraints, though relevant, are alleviated by a well-established infrastructure with continuous energy supply. However, NTNs, particularly satellite-based deployments, operate under stringent power limitations due to the restricted energy resources available on satellites. This fundamental constraint underscores the necessity for intelligent and adaptive energy management techniques to ensure optimal utilization of available power without compromising network performance. One crucial aspect of NTN-O-RAN integration is the functional split between the CU and DU, which relies on the Space-to-Ground link,  implemented via the feeder link. While an initial static design can define an optimal functional split, the dynamic nature of NTNs requires continuous adaptation due to frequent modifications in network structure, topology, demands and resource availability \cite{rodrigues2023hybrid}, \cite{zhu2022delay}. Addressing this challenge necessitates the incorporation of near-real-time Radio Intelligent Controllers (RICs) based on O-RAN principles. These controllers collect real-time network status data, analyze operational parameters, and dynamically adjust the functional split to enhance efficiency. The primary goal of this optimization process is to minimize energy consumption within the satellite payload while maintaining optimal network performance. Key factors considered in this process include traffic demand fluctuations, payload processing capabilities, available energy resources, and the quality of the CU-DU feeder link \cite{campana2023ran}.

Several research efforts have explored RAN functional split strategies mainly in  terrestrial network architectures. The work in \cite{small2016small} discusses various RAN functional split options in accordance with 3GPP guidelines, outlining potential implementations. Similarly, \cite{larsen2018survey} examines the technical requirements and constraints associated with different 3GPP-specified functional splits. Additionally, studies such as \cite{murti2020optimal} and \cite{amiri2023energy} focus on optimizing virtualized network function placement and traffic routing in terrestrial networks, leveraging machine learning-based approaches for efficient resource allocation. More recently, \cite{rihan2023ran} investigates feasible functional split strategies for integrated TN-NTN systems, although a comparative analysis of these options has not been comprehensively explored in the literature. There are limited works introducing potential functional split approaches in NTN systems\cite{khouli2023functional}. However, given the complexity of integrating NTNs with O-RAN architectures, further research is required to evaluate efficient functional split strategies, ensuring optimal energy efficiency while maintaining robust network performance in space-based deployments.
\begin{table*}[htbp]
\vspace{10pt}
\caption{Functional split options between  CU and DU in O-RAN  with respect to the RU traffic load of $\lambda_{RU}$ Mbps }
    \centering
    \begin{tabular}{c|c|c|c|c} 
Split Option $(o)$ & Functions in DU & Functions in CU & Latency (ms) & Peak Traffic (Mbps) \\
\hline 0 &   PHY, RLC, MAC, PDCP & None  & 10 & $\lambda_{RU}$ \\
1 &  PHY, MAC, RLC &  PDCP & 1.5-10 & $\lambda_{RU}$ \\
2 &  PHY, MAC, Low-RLC &  High-RLC, PDCP & 1.5-10 & $\lambda_{RU}$\\
3 & PHY, MAC &  RLC, PDCP & 0.1 & $\lambda_{RU}$\\
4 & PHY, Low-MAC &  High-MAC, RLC, PDCP & 0.1 & $1.02 \lambda_{RU}{+}1.5$\\
5 & PHY &  , MAC, RLC, PDCP & 0.25 & $1.02 \lambda_{RU}{+}1.5$\\
6 & None &  PHY, MAC, RLC, PDCP & 0.25 & 2500\\
\end{tabular}
    \label{tab of spl}
\end{table*}

Building upon the aforementioned considerations, in \cite{shahabi2024energy}, we introduce two potential platforms for creating non-terrestrial O-RAN using either Low Earth Orbit (LEO) satellite-aided RAN or HAPS-aided RAN. This paper delves into the NTN architectures specifically designed in accordance with O-RAN principles to fully utilize network disaggregation options and RAN functional split by considering any combination of TN and NTN nodes. A key focus is placed on leveraging near-real-time RICs to introduce an enhanced level of flexibility, enabling seamless adaptability to different NTN-based RAN deployments. This adaptability facilitates a system-aware and proactive approach to optimizing the functional split across various NTN connectivity nodes, including LEO satellites and HAPS as well as the gateway at ground stations. Depending on specific service requirements, such as stringent latency constraints or high-throughput traffic demands, user terminals (UTs) can dynamically establish connections with the most suitable platform to ensure optimal network performance. This paper specifically addresses the challenge of determining the most energy-efficient functional split strategy for O-RAN configurations in NTNs, considering both system constraints and performance optimization. Generall in O-RAN, the RU is connected to the DU which  serves the RU through fronthaul (FH) link.
The DU, on the other hand, is connected to the CU site, which is directly linked to the core network via midhaul (MH) links, for meeting the delay requirements for the RAN split options.

To achieve this, we introduce a novel framework that extends beyond conventional monolithic gNB implementations at all network nodes. In the proposed approach, the CU and DU can be flexibly allocated either onboard NTN nodes—such as LEO satellites or HAPS—or situated at ground-based gateways. The flexibility in functional placement forms the foundation for our analysis, where we investigate the optimal partitioning of network functions between the CU and DU while adhering to critical constraints such as power consumption limitations, latency requirements, varying traffic intensities, and computational load distribution across different NTN and TN nodes.

To address this optimization problem, we propose a deep Q-network (DQN)-based reinforcement learning framework. The proposed approach is capable of operating effectively in highly dynamic environments characterized by fluctuating traffic patterns and latency-sensitive applications. By continuously learning from real-time network conditions, the DQN model
autonomously selects an appropriate disaggregation mode between NTN and TN nodes and determines the most energy-efficient RAN functional
split. 
\section{System Model}
\begin{figure}[b]\label{SYS_ARCH}
    \centering
    \includegraphics[width=\columnwidth]{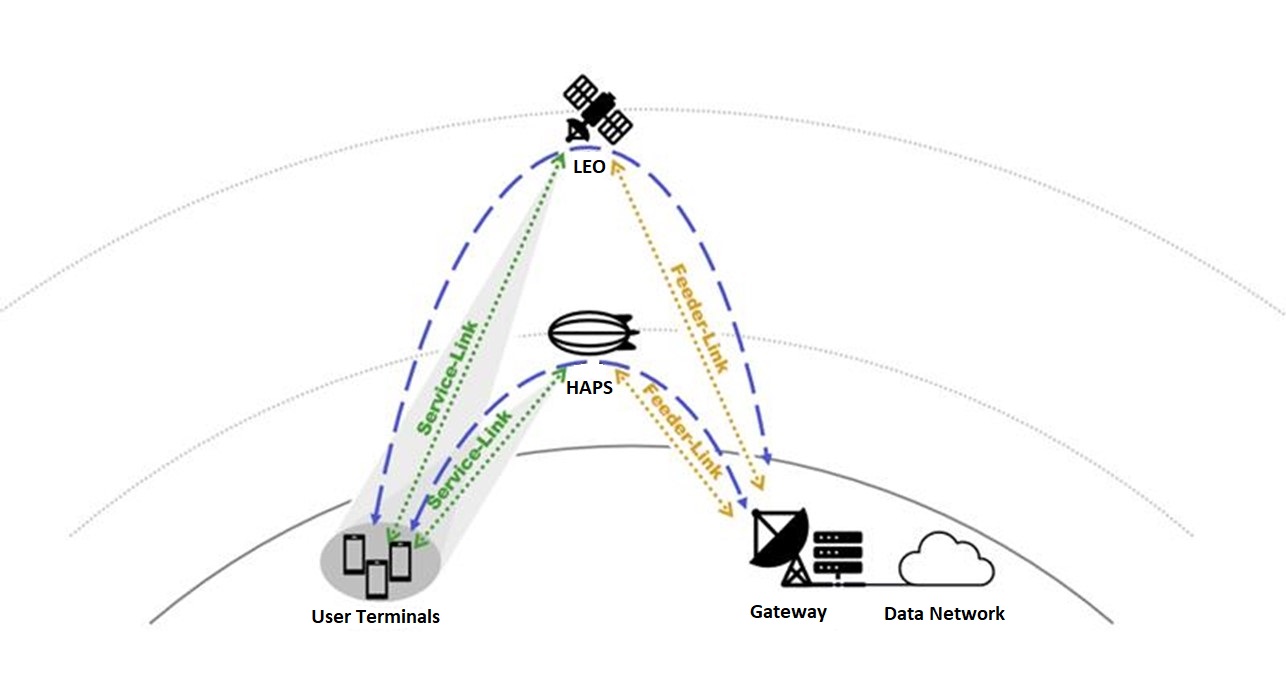}
    \caption{The NTN/TN system model} 
\end{figure}

\begin{figure}[b]\label{splitoptions}
    \centering
    \includegraphics[width=0.8\columnwidth]{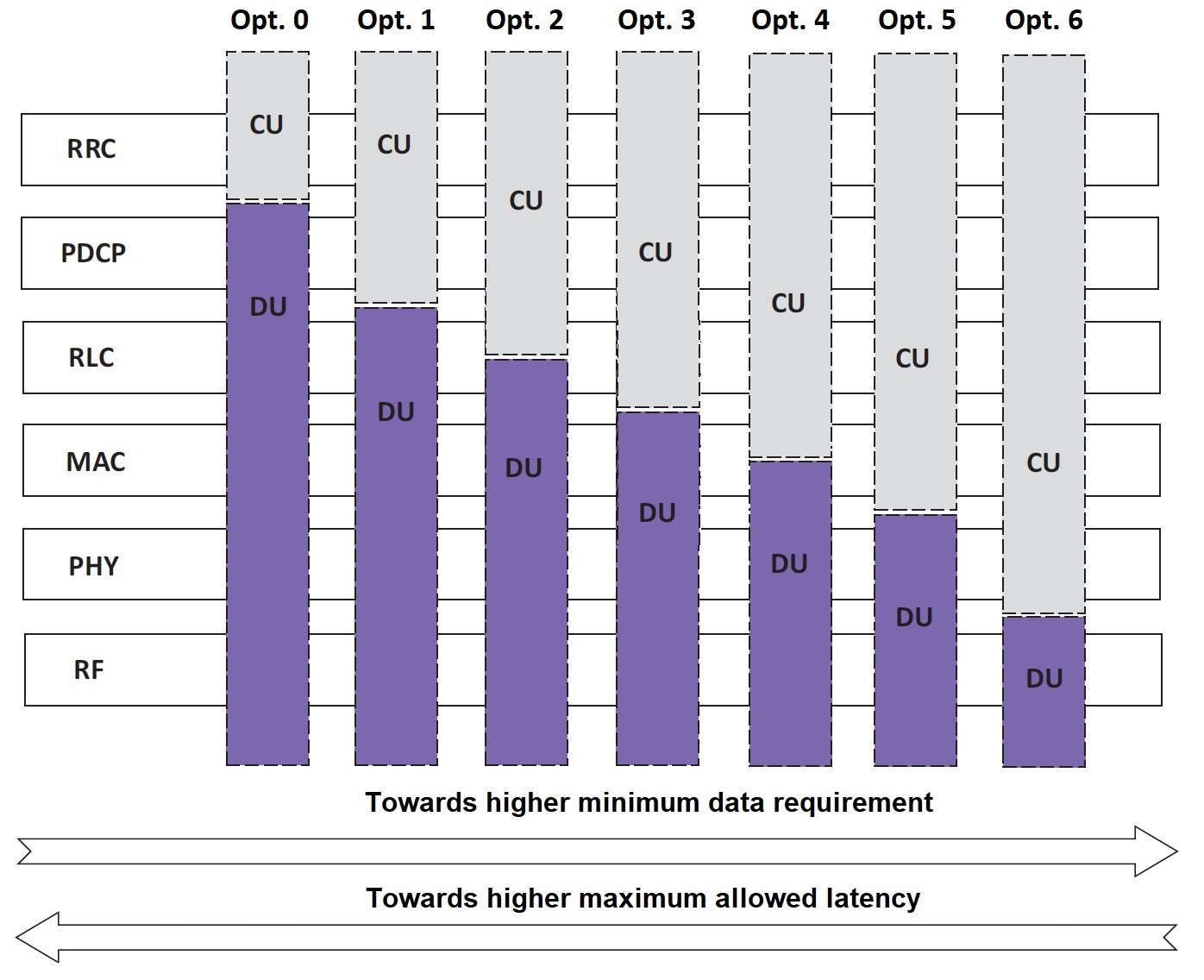}
    \caption{The trade-off between the data rate and the latency for RAN functional split options}
    \label{Overall Arch}
\end{figure}
The system model under investigation, as shown in Fig. 1 enables a number of UTs  to establish connection with NTN nodes, i. e., LEO satellite or HAPS, through the service link. On the other side of the NTN platforms, the NTN nodes are in charge of  communicating with the NTN gateway being connected to the 5G core network via the feeder link. Based on recently defined interfaces  by 3GPP, the NTN nodes (LEO satellites,  HAPSs and gateways) are able to operate as either monolithic gNBs (NR logical nodes) or distributed gNBs with network function disaggregation. 
In case a distributed gNBs is considered, a functional split option should be inevitably selected out of the possible split options in accordance with the network quality of service requirements. 
Based on the 3GPP TR 38.821\cite{3gpp2023solution}, we consider both transparent and  regenerative NTN-based O-RAN architectures being capable of adapting to different NTN-based RANs  that meet the traffic load and latency requirements of the UTs, in which the gNB-RU are located on the NTN nodes and  the gNB-CU and the gNB-DU can be placed on on either the ground station or the NTN nodes depending on the selected disaggregation option to improve the network performance in terms of adapting to traffic demands and energy efficiency requirements\cite{bonafini2022analytical,wang2022seamless,cui2022space}. As a consequence, parts of processing tasks, which are basically performed by the gNB at the ground station or gateway, can be imposed on the NTN nodes.

Generally, the functionality of a gNB  is broken down into a series of functions, whose distributions are determined by functional split options. Accordingly we utilize RAN function split options introduced in \cite{3GPP1} and \cite{3GPP2} as shown in Table I. It is assumed that for each time step $n$, a split option $o\in\mathcal{O}$ is selected from the possible  options in Table. I, where $\mathcal{O}=\{0,1,2,\dots, 6\}$ is the set of possible RAN functional split options. 
\begin{table*}[htbp]
\caption{The computational loads of the CU node and the the DU node for different split options}
    \centering
    \begin{tabular}{c|c|c} 
Split option (o) & $COMP_{DU,c}^{o}$ & $COMP_{CU,c}^{o}$  \\
\hline 0 & $COMP_{PHY}{+}COMP_{MAC}{+}COMP_{RLC}{+}COMP_{PDCP}$ & $0$ \\
1 & $COMP_{PHY}{+}COMP_{MAC}{+}COMP_{RLC}$ & $COMP_{PDCP}$ \\ 
2 & $COMP_{PHY}{+}COMP_{MAC}{+}COMP_{Low-RLC}$ & $COMP_{High-RLC}{+}COMP_{PDCP}$ \\
3 & $COMP_{PHY}{+}COMP_{MAC}$ & $COMP_{RLC}{+}COMP_{PDCP}$ \\
4 & $COMP_{PHY}{+}COMP_{Low-MAC}$ & $COMP_{High-MAC}{+}COMP_{RLC}{+}COMP_{PDCP}$
\\
5 & $COMP_{PHY}$ & $COMP_{MAC}{+}COMP_{RLC}{+}COMP_{PDCP}$\\
6 & $0$ & $COMP_{PHY}{+}COMP_{RLC}{+}COMP_{MAC}{+}COMP_{PDCP}$
\end{tabular}
    
    \label{COMP_SAT,GAT}
\end{table*}
Accordingly, the traffic flows are aggregated among RUs, DUs and CUs. Consequently, a sequence of network functions $\{f0 \rightarrow f1 \rightarrow\dots\rightarrow f6\}$ should be taken into account, in which $f0$ is placed in RU driven by delay stringent requirements, while the other functions might be placed in either CU or DU as virtual machines (VMs). In this sequence, $f1$ denotes the PHY functions in terms of cyclic de-prefix and FFT, resource mapping, QAM, antenna de-mapping and FEC. $f2$ and $f3$ represent the low-MAC and high-MAC functions while $f4$ and $f5$ stand for the low-RLC and high-RLC functions, respectively.  Finally,  $f6$ is the PDCP function.

While  centralization results in enhanced resource management efficiency as well as lower DU complexity and related expenses, it necessitates stricter demands on data rates and latency, as depicted in Fig. 2. In other words, centralizing functions increases  the data load that should  be imposed on the CUs. 

\section{Problem Formulation}
Prior to the RAN functional split optimization, we should be able to calculate the key factors affecting the optimal split option, i. e., power consumption and latency. Therefore, in the following, we obtain these parameters. 
In order to distinguish among potential split options, a binary  variable $x_{c,d}^{o} \in\{0,1\}$ is introduced to define deploying split $o \in \mathcal{O}$ for the NTN-based O-RAN network made up of the CU at  node $c$ and the DU at  node $d$ in each time slot, where $c,d\in\mathbb{N}$ and $\mathbb{N}=\{GAT,SAT,HAP\}$ is the set of the possible NTN nodes, whose entries stand for the gateway, the LEO satellite and the HAPS, respectively. In each time step, it is assumed that the UTs are connected to only one NTN node, while the O-RAN network selects one split option out of the possible options. We consider the following five options for splitting the CU and the DU:
\begin{itemize}
    \item Disaggregation option 1; monolithic gNB at the gateway.
    \item Disaggregation option 2; CU at the gateway and the DU at the LEO satellite.
    \item Disaggregation option 3; monolithic gNB at the LEO satellite.
    \item Disaggregation option 4; CU at the gateway and the DU at the HAPS.
    \item Disaggregation option 5; monolithic gNB at the HAPS.
\end{itemize}
Taking above assumptions into consideration, we have

\begin{equation}\label{binarycond1}
    \sum_{o \in \mathcal{O}} x_{c,d}^{o}=7 \quad \forall c=d, \ c=d \in \mathbb{N},
\end{equation}
\begin{equation}\label{binarycond2}
    \sum_{o \in \mathcal{O}} x_{c,d}^{o}=1, \quad \forall c\neq d, \ c=d \in \mathbb{N},
\end{equation}
\begin{equation}\label{binarycond3}
    \sum_{\substack{d \in \mathbb{N}\\d\neq c}} x_{c,d}^{o}=1, \quad \forall o\in \mathcal{O}, \ c=GAT,
\end{equation}
\begin{equation}\label{binarycond4}
    \sum_{\substack{c \in \mathbb{N}\\c\neq d}} x_{c,d}^{o}=0, \quad \forall o\in \mathcal{O}, \ d=GAT.
\end{equation}
where (\ref{binarycond1}) implies the monolithic gNB options that do not impose any disaggregation and keep all functions centralized in the gNB. In addition, (\ref{binarycond2}) guarantees that only one split option from $\mathcal{O}$ is selected for disaggregated gNBs. Also, disaggregation options 2 and 4 are characterized by (\ref{binarycond3}). It should be noted that for a disaggregated gNB, as shown in (\ref{binarycond4}), when CU is placed on the LEO satellite or HAPS, DU is not allowed to be placed on the gateway and such a disaggregation is not valid according to the disaggregation options discussed earlier. below we calculate the power consumption and latency that entail each aggregation option and the split option.
\subsection{Power Consumption Calculation}
The total power consumption of the RAN functions for the split option $o\in\mathcal{O}$ can be obtained as the sum of the processing power $P_P^{o}$ and transmission power $P_T^{o}$  for that split option as follows
\begin{equation}\label{Total Power}
    P_{Total}^{o}=P_{P}^{o}+P_{T}^{o}.
\end{equation}
in which the processing power consumption is calculated as
\begin{equation}
    P_{P}^{o}=\sum_{c \in \mathcal{N}}\sum_{d \in \mathcal{N}}\left(P_{P-CU,c}^{o}+P_{P-DU,d}^{o}\right)x_{c,d}^{o}.
\end{equation}
The values of $P_{P-CU,c}^{o}, \ c\in\mathbb{N}$  and $P_{P-DU,c}^{o}, \ d\in\mathbb{N}$ denote the processing powers of the CU node $c$ and the DU node $d$  for the split option $o$, respectively, which  are calculated as follows
\begin{equation}\label{P_P_SAT}
    P_{P-CU,c}^{o}=P_{I,c}(2-x_{c,c}^{o})/2+EPO_{c}.COMP_{CU,c}^{o},
\end{equation}
\begin{equation}\label{P_P_SAT2}
    P_{P-DU,d}^{o}=P_{I,d}(2-x_{d,d}^{o})/2+EPO_{d}.COMP_{DU,d}^{o},
\end{equation}
in which $P_{I,i}, \ i\in\mathbb{N}$ is the idle power of  processors/servers at the processing node $i$, $EPO_{i}$ denotes the energy per operation for that node and  $COMP_{CU,c}^{o}$  and $COMP_{DU,d}^{o}$ are the  computational loads in  split option $o$ for  CU node $c$    and  DU node $d$, respectively, which  are given in the TABLE \ref{COMP_SAT,GAT}. The weights $(2-x_{c,c}^{o})/2$ and $(2-x_{d,d}^{o})/2$ only portray the effect of monolithic gNBs on the idle power where we merge the idle powers of the CU and the DU, and would be ineffective for disaggregated modes $(c\neq d)$ as they would be equal to one.

On the other hand, the transmission power $P_T^{o}$ in (\ref{Total Power}) for the split option $o$ is calculated by
\begin{equation}
    P_{T}^{o}=\sum_{\substack{d \in \mathcal{N}\\ d\neq c}}\sum_{c \in \mathcal{N}}\frac{p_{d,c}}{C_{d,c}}TRA^{o}x_{c,d}^{o},
\end{equation}
in which, $C_{d,c}$ is the maximum link capacity between   node $d$ and  node $c$, $p_{d,c}$ denotes the power consumed by  node-$d$/node-$c$ to transmit towards each other at the maximum link capacity and $TRA^{o}$ is the traffic demand of  RAN split option $o$ on the feeder link (the last column of TABLE. I) which is a function of the RU traffic $\lambda_{RU}$.
\subsection{Latency Calculation}
The latency caused by RAN functional split mainly stems from the propagation of electromagnetic waves which is calculated by
\begin{equation}
    L_{Total}^{o}=\sum_{\substack{d \in \mathcal{N}\\ d\neq c}}\sum_{c \in \mathcal{N}}x_{c,d}^{o}\left(\frac{d_{c,d}}{C}\right),
\end{equation}
in which $d_{c,d}, \ \forall d, c \in \mathbb{N}$ denotes the distance between node $c$ and  node $d$ while $C$ represents the speed of the light.
\subsection{RAN functional Split Optimization Problem}
The objective for the optimization of the RAN functional split option is to minimize the total power consumption while meeting the latency and traffic requirements. Hence, the resultant optimization problem is given as follows

\begin{equation}\label{main opt1}
\begin{array}{cl}
\min\limits_{\substack{x_{c,d}^{o} \\ c, d\in\mathbb{N} \\ o\in\mathcal{O}}}& P_{Total}^{o}\\ 
&    \\
\st 
& \text{condition }(\ref{binarycond1}),\\
&     \\
& \text{condition }(\ref{binarycond2}), \\
&   \\
& \text{condition }(\ref{binarycond3}), \\
&   \\
& \text{condition }(\ref{binarycond4}), \\
&   \\
& L_{Total}^{o}(1-x_{c,c}^o)\leq L^{o}, \ \forall c \in \mathbb{N}, \ \forall o \in \mathcal{O},\\
&    \\
& TRA^{o}\leq x_{c,d}^{o}C_{d,c}, \ \forall c,d \in \mathbb{N}, \ \forall o \in \mathcal{O},\\
&    \\
& x_{c,d}^{o}\left(COMP_{CU,c}^{o}+COMP_{DU,d}^{o}\right)\leq COMP_{max,c}\\&+COMP_{max,d}, \ \forall c,d \in \mathbb{N}, \ \forall o \in \mathcal{O}, \\

\end{array}
\end{equation}
in which $L^{o}$ stands for the latency requirement for the split option $o$ (in Table. \ref{tab of spl}). 
Also, $COMP_{max,c}$ and $COMP_{max,d}$ are the maximum computational capacity of the server at the  node $c$ and the node $d$ in GOPS, respectively.
While the first four constraints refer to  conditions earlier mentioned in eq. (\ref{binarycond1}) - (\ref{binarycond4}), the fifth constraint ensures that
the latency between the NTN nodes  does not exceed the maximum tolerable delay for the selected split option, and the sixth constraint ensures that the traffic resulting from the functional split  does not exceed the service link capacity. Also, the seventh  constraint ensures that the computational loads is below the computational capacity of the NTN nodes. To address the optimization problem (\ref{main opt1}) in a real-time manner, we employ  deep reinforcement learning with particular focus on the DQN method which is discussed in the next section.
\section{DQN-based method}
To effectively adapt to dynamic environments and enable efficient decision-making in solving optimization problem (\ref{main opt1}), we propose leveraging a deep reinforcement learning framework based on a DQN. This approach is well-suited for capturing the network’s dynamic behavior in response to fluctuating traffic loads and varying latency constraints in real time. As a reinforcement learning algorithm, DQN is particularly effective in decision-making within dynamic and discrete environments, where both the state and action spaces consist of a finite set of possible values.
The core mechanism of DQN relies on a Q-neural network (QNN) that approximates the optimal action-value function, thereby allowing agents to iteratively refine their decision-making strategies. By interacting with the environment through a sequence of states, actions, and rewards, agents learn to take actions that maximize long-term cumulative rewards. At each step, an action is chosen based on the current state of the system, with the objective of improving performance over successive iterations. The received rewards act as crucial feedback signals, reinforcing beneficial actions while discouraging suboptimal decisions, thus enabling continuous adaptation to the evolving network conditions.
Given the importance of learning an effective control policy, a well-defined set of actions and an appropriate reward function are essential prerequisites for training the DQN model. The reward structure should be carefully designed to guide the system trajectory toward desirable outcomes, ensuring stability and efficiency. Furthermore, constructing a suitable neural network architecture for training is equally critical, as it influences the model’s ability to generalize across different network scenarios and effectively optimize resource allocation in the presence of unpredictable variations.
\subsection{State Set}
Definition of states in DQN should comprehensively describe optimization environment. So, in each time step $n$, the state set is defined as follows
\begin{align}\label{state set}
    s[n]{=}&\Big\{o[n],c[n],d[n],TRA^{o[n]},L^{o[n]},\lambda_{RU}[n],P_{Total}^{o[n]},
    \nonumber\\
    &L_{Total}^{o[n]},C_{c[n],d[n]},
    COMP_{CU,c[n]}^{o[n]}, COMP_{DU,d[n]}^{o[n]},
    \nonumber\\
    &
    COMP_{max,c[n]}, COMP_{max,d[n]}\Big\},
\end{align}
where $\lambda_{RU}[n]$, $o[n]$, $c[n]$ and $d[n]$  denote the RU traffic load, the selected split option the selected CU node and the selected DU node in the time step $n$, respectively.
\subsection{Action Set}
Considering initial assumptions are made for the split option and network disaggregation level, i. e., monolithic gNBs or disaggregated gNBs, the action set at time step $n$ consists of two separate actions:

\textbf{Action 1 (For placing CU and DU)}: Assuming a selected CU/DU placement in the previous time step, i. e., LEO satellite, or HAPS or gateway, at time step $n$, Action 1 denoted by ${a_1} [n]$ is selecting one option from the following options:
\begin{itemize}
    \item Monolithic gNodeB at gateway (both CU-DU at gateway).
    \item CU at gateway and DU at LEO.
    \item Monthilic gNodeB at LEO (both CU-DU at LEO)
    \item CU at gateway and DU at HAP
    \item Monthilic gNodeB at HAP (both CU-DU at HAP)
    \item Keep the previous placement unchanged
\end{itemize}

\textbf{Action 2 (for selecting a split option)}: Given a selected split option at the previous time step, an action represnted by ${a_2} [n],$ is taken at time step $n$ out of the following three options (based on the latency requirement and traffic flow of the selected split option):
\begin{itemize}
    \item Moving RAN split point towards the first-upper split point
    \item Moving RAN split point towards the first-lower split point
    \item No action is taken,
\end{itemize}
\textcolor{black}{in which the split point is where the network functions is splitted to DU functions and the CU functions (Fig. 1).} The total action in the $n$-th time step is defined as 
\begin{equation}\label{action set}
    a[n]=\{{a_1} [n], {a_2} [n]\}.
\end{equation}
The reason behind such an action definition is twofold. First, this definition can simply show the impacts of the split point and split option on the traffic flow and latency requirement, i. e., based on the Fig. 2, moving split point towards upper-layer points at the protocol stack can usually result in lower traffic demand but higher latency, and vice versa. Second, defining this action with only three options leads to a small action space size.

Assuming six options for Action 1 and three options for Action 2, the total action space is
18 which is a reasonable size for a DQN method. 

\subsection{Reward}

To fully characterize network performance and provide a correct trajectory for the DQN method, the reward in the time step $n$ is defined as follows
\begin{equation}\label{reward set}
    r[n]=\sum_{j=1}^{5}\nu_{j}R_{j}[n],
\end{equation}
where
\begin{equation}\label{rew1}
    R_{1}[n]=\begin{cases}
        +1,& L_{Total}^{o[n]}(1-x_{c[n],c[n]}^{o[n]})\leq L^{o[n]}, \\
        -1,&  L_{Total}^{o[n]}(1-x_{c[n],c[n]}^{o[n]})> L^{o[n]},
    \end{cases}
\end{equation}
\begin{equation}
    R_{2}[n]=\begin{cases}
        +1,& TRA^{o[n]}\leq x_{c[n],d[n]}^{o[n]}C_{d[n],c[n]}, \\
        -1,& TRA^{o[n]}> x_{c[n],d[n]}^{o[n]}C_{d[n],c[n]},
    \end{cases}
\end{equation}
\begin{equation}
    R_{3}[n]{=}\begin{cases}
        {+}1,& x_{c[n],d[n]}^{o[n]}\left(COMP_{CU,c[n]}^{o[n]}{+}COMP_{DU,d[n]}^{o[n]}\right)\\&{\leq} COMP_{max,c[n]}{+}COMP_{max,d[n]}, \\ & \\
        {-}1,& x_{c[n],d[n]}^{o[n]}\left(COMP_{CU,c[n]}^{o[n]}{+}COMP_{DU,d[n]}^{o[n]}\right)\\& {>}COMP_{max,c[n]}{+}COMP_{max,d[n]},
    \end{cases}
\end{equation}
\begin{equation}\label{overhead reward for CU/DU Placement}
    R_{4}[n]=\left\{\begin{array}{ll}
        -1, & \text{New CU/DU placement in  time step $n$,}  \\
         0, & \text{Keeping the previous placement},
    \end{array}\right.
\end{equation}
\begin{equation}\label{overhead reward for RAN functional split}
    R_{5}[n]=\left\{\begin{array}{ll}
        -1, & \text{Moving the RAN split point in time step $n$,}  \\
         0, & \text{Keeping the previous RAN split point},
    \end{array}\right.
\end{equation}
while $\nu_{j}, \ j\in\{1,2,5\}$ are penalty positive-value coefficients for $R_{j}[n]$ that are set during the training process with appropriate weights.
The values of $R_{1}[n]$, $R_{2}[n]$ and $R_{3}[n]$ reflect the constraints coming from latency requirement, traffic load and computational capacity, respectively. Also, $R_{4}[n]$ and $R_{5}[n]$ imply the cost induced by taking an action in the $n$-th time step for CU/DU placement and RAN functional split, respectively, i.e., each action results in a negative reward irrespective to other  reward parameters $R_{1}[n]$ to $R_{3}[n]$. These two parameters may result in faster algorithm convergence by avoiding unnecessary changes in CU/DU placement and RAN functional split. In other words, by changing these two architecture parameters, negative rewards are obtained unless these changes cause sufficient positive values for reward parameters $R_{1}[n]$ to $R_{3}[n]$ so that the total reward in (\ref{reward set}) becomes positive.
It is noteworthy that selecting options for Action 1 and Action 2 results in setting a specific value for $x_{c,d}^{o}$ in the time step the action is taken. The obtained value for this variable will be assessed in each time step based on the reward defined above.

\begin{algorithm}[t] 
\caption{DQN-based Method}
\begin{algorithmic}\label{Alg1}
\REQUIRE $s_{0}, \varepsilon, \mu, \rho, N_{B}, T$
\STATE $\mb{Initialize}$ the experience memory \STATE $\mb{Initialize}$ the parameter of QNN as $\bs{\omega}$
\STATE $\mb{Initialize}$ the parameter of target QNN $\hat{\bs{\omega}}=\bs{\omega}$

\FOR {$t=0,1,2, \cdots$ in DQN}  
\STATE Input $s_{t}$ to QNN and output $Q=\left\{q\left(s_{t}, a, \bs{\omega}\right)|a \in \mathbb{A}\right\}$
\STATE Select action $a_{t}$ from $Q$ using $\varepsilon$ -greedy algorithm 
\STATE Receive $r_{t+1}$ 
\STATE Calculate $s_{t+1}$ from $s_{t}, a_{t}$ and $r_{t+1}$
\STATE Store $\left(s_{t}, a_{t}, r_{t+1}, s_{t+1}\right)$ to the experience memory
 \STATE $\mb{if}$  $t/T==0$ $\mb{then}$ $F=1$ $\mb{else}$ $F=0$ 

\STATE Randomly select $N_{B}$ samples from the memory as $B$ 
\FOR {each sample $e=\left(s, a, r, s^{\prime}\right)$ in $B$} 
\STATE Calculate $x_{(r,s')}^{\text{target}}=r+\mu \max _{a^{\prime}} q\left(s^{\prime},a^{\prime};\hat{\bs{\omega}}\right)$
\ENDFOR 
\STATE $\bs{\omega}\leftarrow\bs{\omega}+\frac{\rho}{N_{B}}\sum_{b\in B}\left[x_{(r,s')}^{\text{target}}-q(s,a;\bs{\omega})\right]\nabla q(s,a; \bs{\omega})$
\IF{$F==1$}  
\STATE Update $\hat{\bs{\omega}}$ in target QNN as  $\hat{\bs{\omega}}=\bs{\omega}$
\ENDIF

\ENDFOR

\end{algorithmic}
\end{algorithm}

\subsection{DQN Training Structure}
  Now, we formulate optimization problem  (\ref{main opt1}) as a Markov decision process (MDP), represented by the tuple ($\mathbb{S} , \mathbb{A} , \mathcal{P}, \mathbb{R}$), in which $\mathbb{S}$, $\mathbb{A}$ and $\mathbb{R}$ represent the state space, the action space and the reward space, whose entries are defined earlier in (\ref{state set}), (\ref{action set}) and (\ref{reward set}) , and $\mathcal{P}$ stands for the probability distribution of transition form one state to another. 
  Based on reinforcement learning,  reward function $R$ maps the state space and action space to the reward space, i. e., $R:\mathbb{S}\times\mathbb{A}\rightarrow\mathbb{R}$. Accordingly,  the $t$-th episode return is defined as 
      $R_t=\sum_{i=1}^T\mu^{i-t}R(a[i],s[i])$, where $T$ is the episode length representing the number of time steps in each episode, $\mu\in[0,1]$ stands for the discount factor of the future rewards, while $a[i]\in\mathbb{A}$  and $s[i]\in\mathbb{S}$.
  By taking an action,  the agent transitions to a new state and  gains a reward.
  Utilizing Q-learning \cite{Sutton2018}, an efficient policy is adopted to estimate the state-action value function which is called Q-function, for maximizing the future reward. 
  The value function is defined as the expected return over all episodes, when starting from  state $s$ and performing  action $a$ by following the policy $\pi:\mathbb{S}\rightarrow\mathbb{A}$ as 
  \begin{equation}
      Q^{\pi}(s,a)=\mathbb{E}[R_t|s_t=s,a_t=a,\pi],
  \end{equation}
  in which $s_t$ an $a_t$ denote the state and action at the beginning of the $t$-th episode, respectively.
  Now,  we can rewrite the objective to maximize the expectation of
the value function over all episodes.
  Employing Bellman equation \cite{Sutton2018}, the optimal value function is obtained by $Q^{*}(s,a)=\max_{\pi}Q^{\pi}(s,a)$ and the optimal policy follows $\pi^*(s)=\arg\max_{a\in \mathbb{A}} Q^{*}(s,a)$.

 Furthermore, in case of a large number of states, DQN is utilized as a candidate solution for systems that suffer from infeasible generalization in unobserved states. 
  To the best of our knowledge, DRL accounts for using neural networks to approximate the Q-function.
  The resultant structure referred to as QNN leads to the following approximation for the action-value function $q(s,a,\boldsymbol{\omega})\approx Q^{*}(s,a)$, in which $\boldsymbol{\omega}$ refers to some parameters that define the Q-value.
  In a specific episode $t$, where the state is $s_t$ and the QNN weights are $\boldsymbol{\omega}$, the DQN agent takes an action with regards to $a_t=\arg_a \max q(s,a,\boldsymbol{\omega})$ where $q(s,a,\boldsymbol{\omega})$ is the output of QNN for every possible action $a$. Then, the agent receives the reward $r_{t+1}$ and transitions to the state $s_{t+1}$. Thus, the experience set at episode $t$ is $(s_t,a_t,r_{t+1},s_{t+1})$ which is used in training QNN.
  The Q-value $q(s,a,\boldsymbol{\omega})$ is then updated towards the target output for the QNN represented by 
  \begin{equation}\label{target x}
    x^{target}_{r_{t+1},s_{t+1}}=r_{t+1}+\mu \max_{a} q(s_{t+1},a,\boldsymbol{\omega}).
\end{equation}
The update rule in DQN is to find the value of $\boldsymbol{\omega}$ in QNN through a training phase, in which the square loss of $q(s,a,\boldsymbol{\omega})$ is minimized.
  The square loss of $q(s,a,\boldsymbol{\omega})$ is defined as
  \begin{equation}\label{prediction error}
      w(s_t,a_t,r_{t+1},s_{t+1})=(x^{target}_{r_{t+1},s_{t+1}}-q(s_t,a_t,\boldsymbol{\omega}))^2,
  \end{equation}
Furthermore, the values of $\boldsymbol{\omega}$, are updated by a semi-definite gradient scheme used for minimizing (\ref{prediction error}) as
\begin{equation}\label{Omega update}
  \boldsymbol{\omega}\leftarrow \boldsymbol{\omega}+\rho  [x^{target}_{r_{t+1},s_{t+1}}-q(s_{t},a_t,\boldsymbol{\omega})] \bigtriangledown q(s_{t},a_t,\boldsymbol{\omega}),
\end{equation}
$\rho$ represents the learning rate step size. As the QNN weights are iteratively updated, the corresponding target value also undergoes changes. In DQN, a quasi-static target network approach is employed, wherein the target Q-network $q(.)$ in (\ref{target x}) is substituted with $q(s_{t},a_t,\hat{\boldsymbol{\omega}})$. The parameter $\hat{\boldsymbol{\omega}}$ is then periodically refreshed every $T$ time steps, following the update rule $\hat{\boldsymbol{\omega}}=\boldsymbol{\omega}$.

Additionally, to enhance system stability, the experience replay method \cite{ref27TCOM} can be incorporated. Rather than performing QNN training sequentially at each time step using only a single experience, this technique enables batch training by utilizing multiple stored experiences. Specifically, a replay memory of fixed capacity is maintained, where experience tuples $v=(s,a,r,s^{'})$ are recorded at designated time steps. From this stored data, a mini-batch $B$ consisting of $N_{B}$ randomly sampled experiences is selected for training, and the corresponding loss function is computed based on these samples.

By integrating the experience replay technique alongside the quasi-static target network approach, the parameter $\boldsymbol{\omega}$ is updated as follows
\begin{align}
    \boldsymbol{\omega}\leftarrow \boldsymbol{\omega}+\frac{\rho}{N_{B}}\sum_{b\in B} [x^{target}_{r,s^{'}}-q(s,a,\boldsymbol{\omega})] \bigtriangledown q(s,a,\boldsymbol{\omega}),
\end{align}
where $x^{target}_{r,s^{'}}=r+\mu \max_{a^{'}} q(s^{'},a^{'},\hat{\boldsymbol{\omega}})$. The whole DQN-based method is summarized in Algorithm \ref{Alg1}.

\section{Numerical Analysis}
\begin{figure}[t]
    \centering
    \includegraphics[width=0.95\columnwidth]{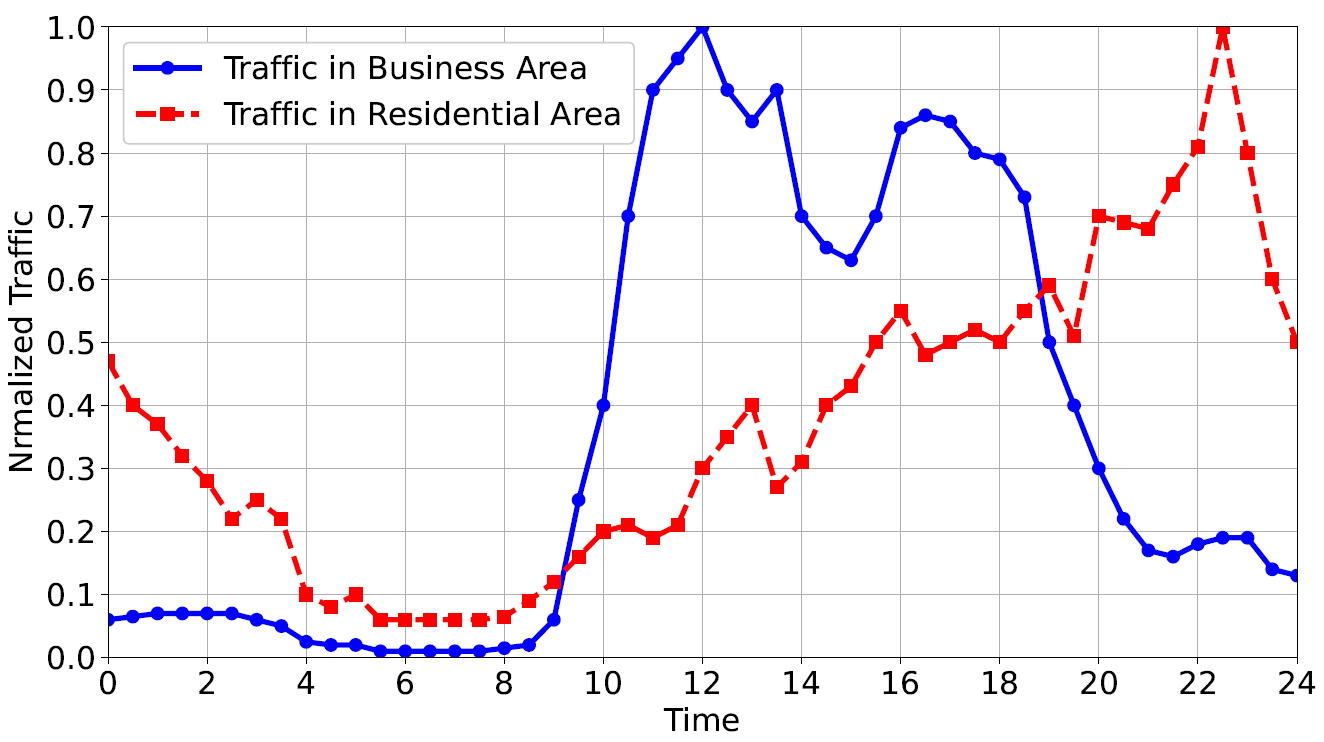}
    \caption{Daily traffic pattern in a residential area and a business area  for weekdays based on measurements in \cite{marsan2013towards}.}
    \label{Traffic Model}
\end{figure}

\begin{figure}
    \centering
    \includegraphics[width=\columnwidth]{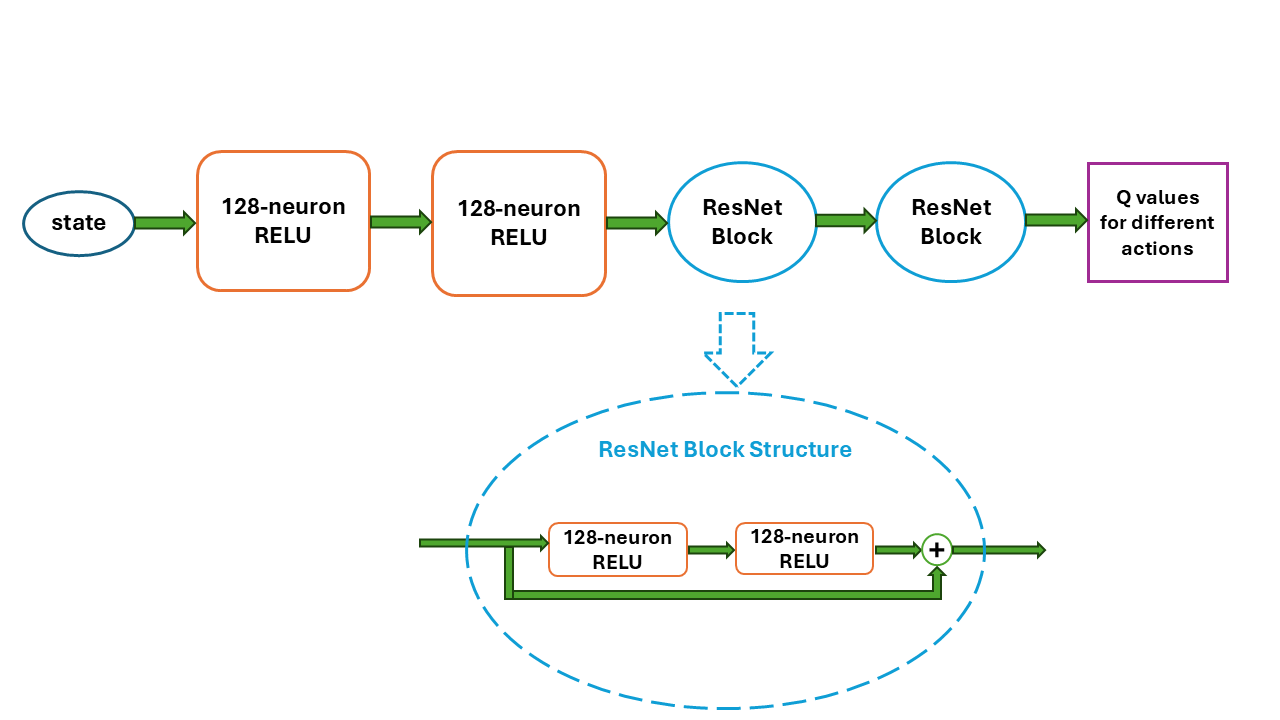}
    \caption{ResNet-aided QNN structure}
    \label{ResNet QNN}
\end{figure}
In this section, we evaluate the performance of the proposed method under varying traffic loads and latency requirements coming from the selected aggregation option and split option according to Table I for downlink transmission over the feeder link. To simulate a dynamic RU traffic flow, denoted as $\lambda_{RU}$, we incorporate the daily traffic patterns outlined in \cite{marsan2013towards}, which represent both business and residential areas on weekdays. These traffic patterns exhibit peak traffic loads of up to 200 Mbps and average traffic loads of 100 Mbps. Fig. \ref{Traffic Model} illustrates the corresponding RU traffic pattern in greater detail, showing the fluctuations throughout a typical weekday cycle. This allows us to model realistic traffic conditions and stress-test the proposed solution under different demand profiles. As the primary metric for performance evaluation, we consider normalized power consumption across the system, allowing for fair comparison across different traffic and load conditions.

The parameters used for the NTN nodes and gateway, are detailed in Table III. These parameters include values for $EPO$s, computational capacities, and idle power levels, which are derived from real-world hardware platforms. Specifically, the Nvidia Jetson TX2 \cite{website_example1}, Nvidia Jetson AGX Xavier \cite{website_example2}, and Nvidia L4 \cite{website_example3} serve as proxies  for the processing units deployed at the HAPS, LEO satellite, and gateway, respectively. These values are selected to ensure an accurate representation of the energy consumption and computational performance in the NTN system, reflecting realistic constraints that would be encountered in an operational environment.

In this work, it is assumed that the energy consumption at the NTN nodes remains within their battery discharge limit. This assumption helps in maintaining the operational integrity of the system. However, it is worth noting that future research will aim to relax this assumption and investigate more challenging scenarios where energy limitations may become a significant factor. This would require the implementation of additional power management strategies to ensure continued service delivery.
\begin{table}[t]
\vspace{15pt}
\caption{Simulation Parameters \cite{3gpp2023solution},\cite{lou2023haps} and \cite{matoussi20205g}}
    \label{tab:my_label}
    \centering
    \begin{tabular}{c|c}
        Parameter & Value (for $i\in\{SAT, HAP, GAT\}$)\\
        \hline
        $COMP_{PHY}$ &  1280 GOPS  \\
        $COMP_{RLC}$ &  50 GOPS  \\
        $COMP_{MAC}$ &  50 GOPS \\
        $COMP_{PDCP}$ &  100 GOPS  \\
        $COMP_{max,i}$ & $SAT$: 32 TOPS, $HAP$: 1.33 TOPS, $GAT$: 485 TOPS   \\
        $d_{i,GAT}$  &  $SAT$: 600 km, $HAP$: 20 km     \\
        $EPO_{i}$  &   $SAT$: 0.625 J/TO, $HAP$: 5,64 J/TO, $GAT$: 0.0742 J/TO  \\

        $P_{I,i}$  & $SAT$: 10 W , $HAP$: 7.5 W, $GAT$: 36 W      \\
        $C_{i,GAT}$ &  $SAT$: 100 Mbps, $HAP$: 10 Gbps \\ 
        $p_{i,GAT}$ &  $SAT$: 35 W, $HAP$: 4 W \\
    \end{tabular}
\end{table}
The DQN framework leverages a deep residual network (ResNet) architecture \cite{he2016deep}, which is composed of six hidden layers, as shown in Fig. \ref{ResNet QNN}. Each hidden layer contains 128 neurons to ensure sufficient representation capacity. For clarity throughout this work, we refer to this QNN, built on a ResNet structure, simply as ResNet. The architecture is organized into segments, where each segment is termed a ResNet block. The neurons in these layers employ ReLU activation functions to introduce non-linearity into the model. The first two hidden layers of this ResNet are fully connected layers, followed by two ResNet blocks, which further enhance learning by adding depth.

Each ResNet block is composed of two consecutive hidden layers with a residual (shortcut) connection that bypasses these two layers and connects the block's input directly to its output. This residual connection mitigates issues like vanishing gradients and allows for better gradient flow during backpropagation, enabling the network to train deeper architectures effectively. To update the QNN's weights, a mini-batch with $N_{B}=32$ samples is randomly selected from an experience replay buffer, which stores the 200 most recent experiences encountered by the agent. The experience replay buffer operates on a first-in-first-out (FIFO) mechanism, where older experiences are discarded as new ones are added, ensuring the buffer remains current. This mini-batch is used to compute the loss function and backpropagate errors.

For optimizing the network, the RMSprop optimization algorithm \cite{goodfellow2016deep} is applied, which adjusts the QNN's weights using the mini-batch gradient descent technique. The use of RMSprop helps to maintain a steady learning rate during training and prevents oscillations in the learning process. Additionally, an exponential decay $\epsilon$-greedy strategy is adopted to guide the exploration process in the deep reinforcement learning (DRL) framework. This technique helps in balancing the trade-off between exploration and exploitation, thus preventing the system from prematurely converging on a sub-optimal policy before sufficient experience is collected. Initially, the exploration rate $\epsilon$ is set to 0.5, encouraging the model to explore various actions. Over time, $\epsilon$ decays by a factor of 0.995 after each time step until a minimum value of 0.0005 is reached, at which point the system prioritizes exploitation over exploration while still maintaining some degree of randomness. In addition, the discount factor $\mu$ and the episode length $T$ are set to 0.9 and 100, respectively.

\begin{figure}[t]
    \centering
    \includegraphics[width=\columnwidth]{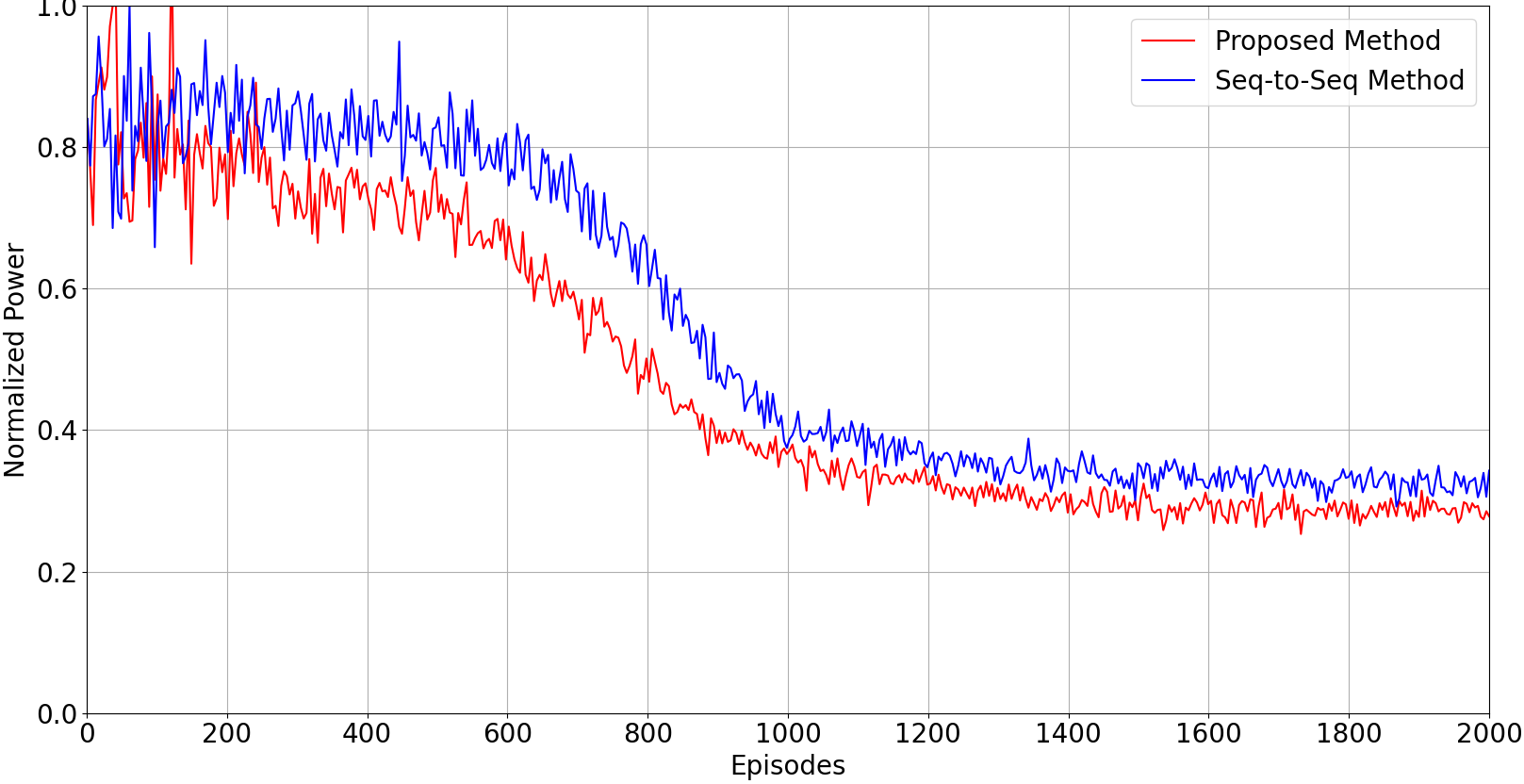}
    \caption{Normalized power vs episodes for a business area}
    \label{NORPoWBus}
\end{figure}

\begin{figure}[t]
    \centering
    \includegraphics[width=\columnwidth]{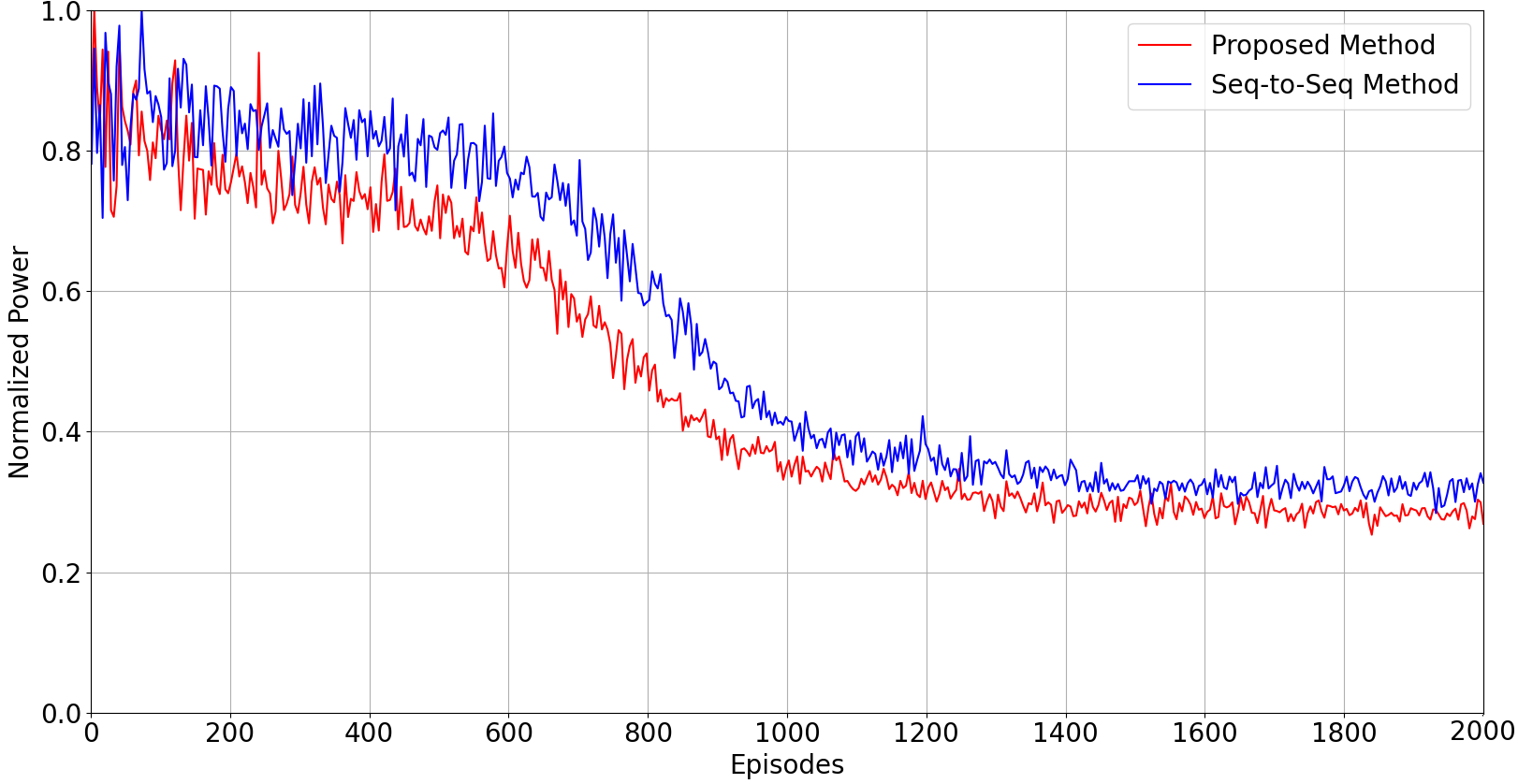}
    \caption{Normalized power vs episodes for a residential area}
    \label{NorPowRes}
\end{figure}
As depicted in Fig. \ref{NORPoWBus}, within a business area, the decision-making process begins with an initial random selection of an action set. This involves choosing a suitable split option along with an appropriate RAN network configuration that adheres to predefined optimization constraints. As the deep Q-network (DQN) agent interacts with the environment, it progressively refines its decision-making strategy by learning from the feedback it receives. Over multiple iterations, the agent systematically adjusts its actions, leveraging reward signals to enhance its policy and move along a more optimal trajectory. In order to have a comparison benchmark, we apply the NTN architecture to the Seq-to-Seq method \cite{amiri2023energy} and extend its split options to those of Table II. Note that the Seq-to-Seq method was aimed for RAN functional split in conventional terrestrial O-RAN with limited split options through optimization using Actor-Critic (A2C) algorithm.  The experimental results highlight that the proposed method not only facilitates a 20$\%$ reduction in normalized power consumption relative to the Seq-to-Seq approach but also exhibits a significantly faster convergence rate. Furthermore, a comparable performance improvement is evident in residential areas, as illustrated in Fig. \ref{NorPowRes}. These findings collectively underscore the effectiveness and efficiency of the proposed approach within the context of NTN-based O-RAN architectures, reinforcing its adaptability to diverse deployment scenarios.

\begin{figure}[t]
    \centering
    \includegraphics[width=\columnwidth]{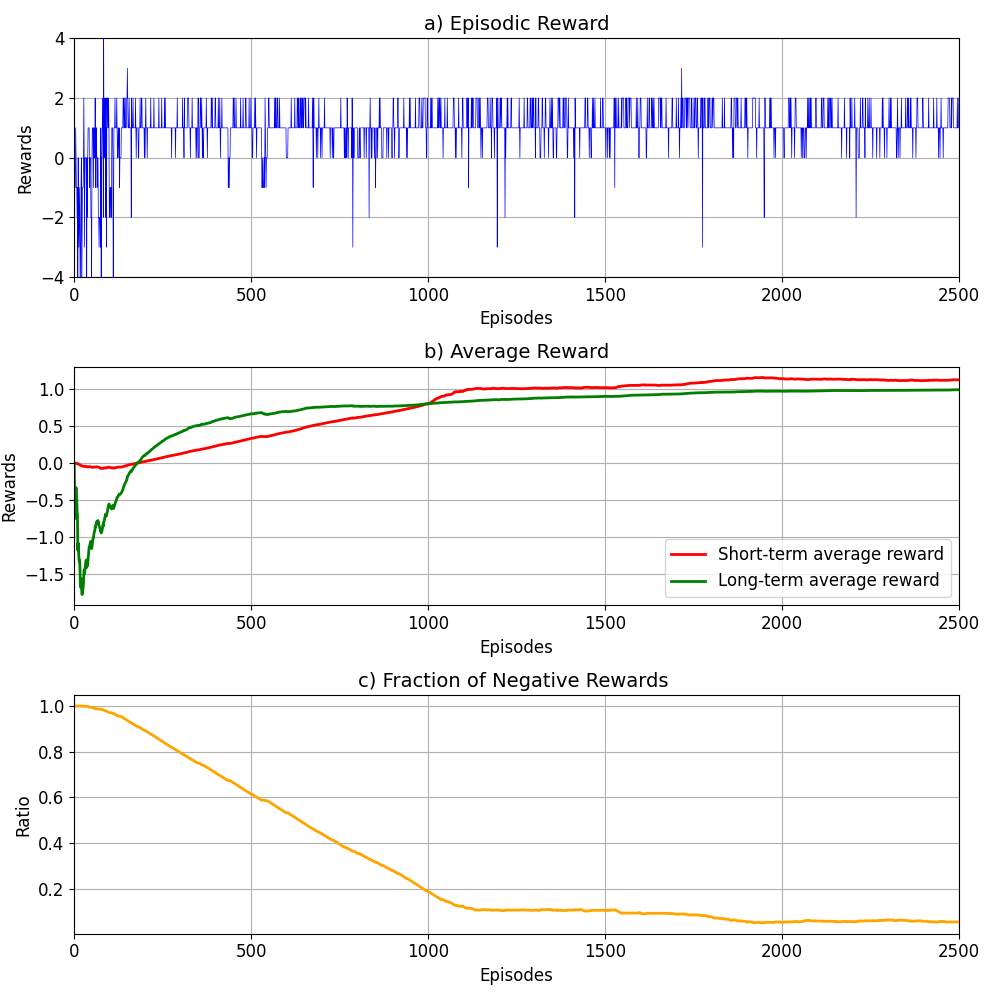}
    \caption{DQN agent reward versus episodes.}
    \label{RewardsPlot}
\end{figure}

In Fig. \ref{RewardsPlot}, the variations of instantaneous reward, short-term reward, and long-term reward over time are illustrated. It is important to highlight that the short-term reward is determined by computing the average of the instantaneous rewards obtained over the most recent 50 episodes. In contrast, the long-term reward is derived by averaging the instantaneous rewards accumulated from the start of the learning process up to the current episode, providing a broader perspective on the overall reward trend.

As depicted in Fig. \ref{RewardsPlot} (a), throughout the learning process, the algorithm encounters both positive and negative instantaneous rewards at different time steps. However, as training progresses and more episodes are completed, the frequency of positive rewards gradually surpasses that of negative ones. This leads to a situation where the majority of instantaneous rewards are clustered within the range [0,2], indicating a favorable learning trajectory. Furthermore, as shown in Fig. \ref{RewardsPlot} (b), after a few episodes, both the short-term and long-term rewards consistently maintain positive values. This observation strongly suggests that the DQN algorithm successfully accumulates positive rewards over time, reinforcing the notion that the learning process follows a stable and effective trajectory. 
Additionally, Fig. \ref{RewardsPlot} (c) presents the reward ratio, which represents the proportion of negative instantaneous rewards relative to the total number of instantaneous rewards obtained. Notably, as the number of time steps increases, the reward ratio steadily declines. Eventually, for sufficiently large time steps, the fraction of negative rewards remains below 0.1, further validating the stability of the learning process. These results confirm that the DQN algorithm successfully converges toward an optimized pilot assignment policy, ensuring efficient decision-making in the long run.

\begin{figure}[t]
    \centering
    \includegraphics[width=\columnwidth]{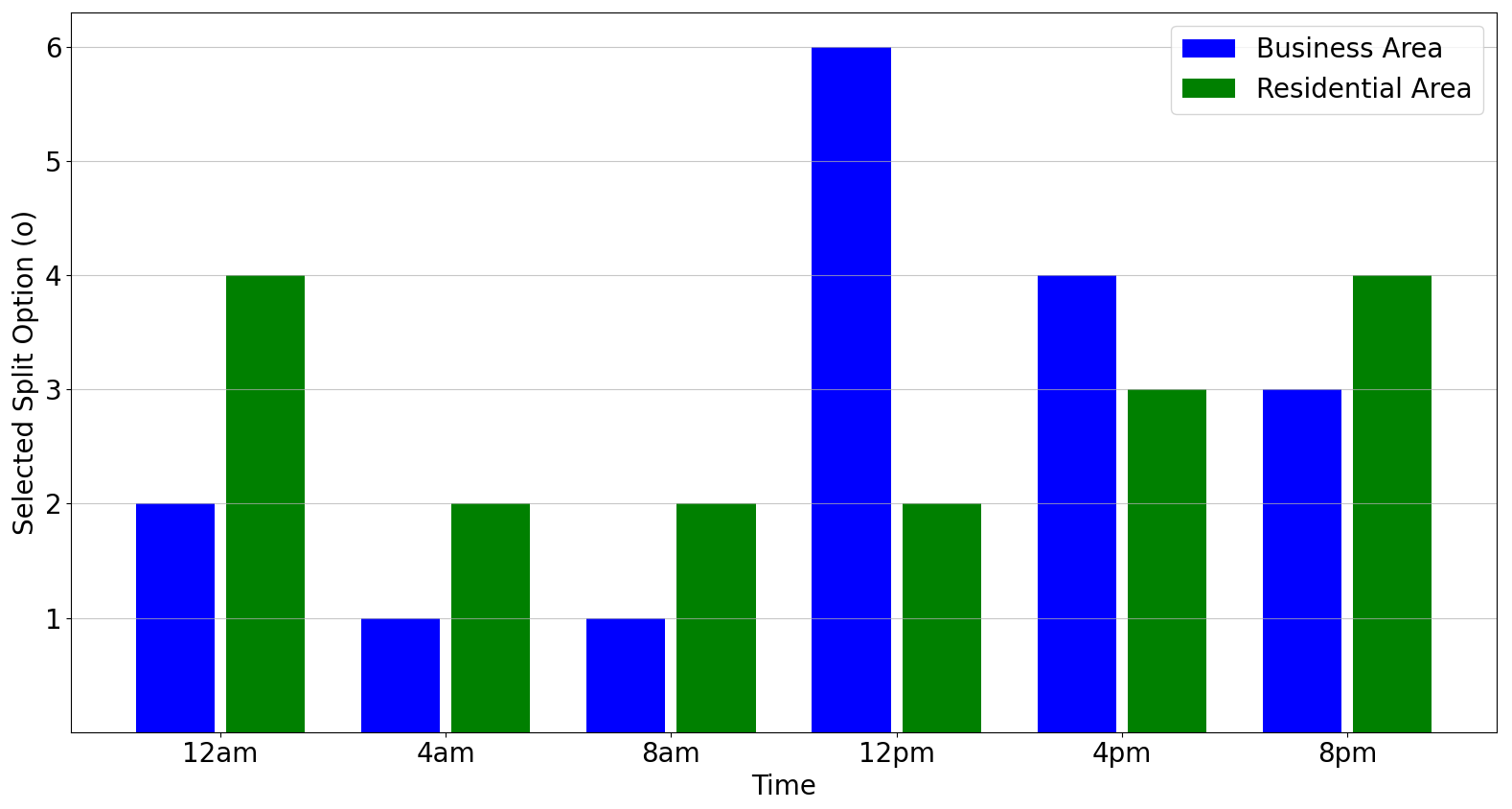}
    \caption{The selected split option for different times of a day}
    \label{daily selected option}
\end{figure}

Fig. \ref{daily selected option} further highlights the adaptability of the proposed framework to dynamic traffic patterns. The figure shows how different split options are selected throughout the day in response to varying traffic conditions. As shown in this figure, in those times of a day when higher traffic flow come from the RU, e. g., 12pm in business area and 8pm in residential area, higher split options are selected. On the other hand, times with lower traffic flow results in lower split options.

\section{Conclusion}
This paper investigated the integration of O-RAN with NTNs, focusing on optimizing the CU-DU functional split for improved energy efficiency. We proposed a DQN-based reinforcement learning framework that dynamically adapts to varying traffic demands, network conditions, and power constraints across platforms like LEO satellites and HAPS. The approach supports different levels of network disaggregation, enabling flexible and efficient resource allocation. The results confirm the capability of the proposed policy to dynamically respond to varying traffic conditions by choosing suitable disaggregation and functional split configurations that align with specific data rate and latency demands.

\bibliographystyle{IEEEtran}
\bibliography{Ref1}

\end{document}